# Grid Cells Are Ubiquitous in Neural Networks

Songlin Li[1]†, Yangdong Deng[2]*†, Zhihua Wang[1]

[1] Department of Microelectronics and Nanoelectronics, Tsinghua University

[2] School of Software, Tsinghua University

* Corresponding author: Email dengyd@tsinghua.edu.cn

† The two authors contributed equally to this work.

## Abstract:

Grid cells are believed to play an important role in both spatial and non-spatial cognition tasks. A recent study observed the emergence of grid cells in an LSTM for path integration. The connection between biological and artificial neural networks underlying the seemingly similarity, as well as the application domain of grid cells in deep neural networks (DNNs), expect further exploration. This work demonstrated that grid cells could be replicated in either pure vision based or vision guided path integration DNNs for navigation under a proper setting of training parameters. We also show that grid-like behaviors arise in feedforward DNNs for non-spatial tasks. Our findings support that the grid coding is an effective representation for both biological and artificial networks.

## Introduction:

It has been long hypothesized that animals use a "cognitive map"[1], namely, an internal knowledge representation to support intelligent behaviors. This proposal was later proved by neuroscience experiments on both spatial[2,3] and non-spatial[4,5] cognition tasks. Grid cells, which exhibit a spatial periodic firing pattern and thus constitute the hexagonal spatial representation as Fyhn et al has observed[2], were discovered in the medial entorhinal cortex (MEC) of rodents[2,3]. Subsequent studies proposed models to characterize the behavior of grid cells in terms of spike-timing correlations[6-8]. The discovery of grid cells as well as the fact that different grid cells forms a multi-scale representation inspired the theory of vector-based navigation[9]. Besides spatial or navigation tasks, researchers also provide experimental proof that grid cells play an important role in the concept space for non-spatial cognition tasks[4,5]. Due to their capability to provide a coding mechanism for abstraction and generalization, grid cells recently attracted the attention of AI researchers. An important discovery is that deep neural networks tuned for navigation tasks reproduce such grid cells[10]. With a proper ensemble coding, a long-short term memory network performing path integration in a square space can replicate hexagonal activation patterns that resemble the neural firing map of grid cells[10]. While providing evidence for the existence of an underlying connection between human brains and deep neural networks (DNNs), the discovery at the same time triggers many open questions.

First, why do both biological networks and DNNs develop a similar hexagonal firing pattern? The significant difference between the underlying mechanisms of the two types of networks makes it hard to tell if the similarity is coincidental or inevitable. Although a theoretically rigorous treatment of the above problem is still open, the results of our work suggest that the similarity is the joint result of optimization and encoding. Human brains are gradually sculptured by the evolution process, while DNNs are explicitly tuned by mathematical optimization algorithms like backpropagation. Previous works demonstrated that memory based DNNs like long short term memory (LSTM) replicated grid cells[10,11] when performing speed-based path integration in navigation tasks. Our results show that the grid firing pattern is a more general behavior that can be replicated in both memory and memory-less networks with and without explicitly considering path integration. Moreover, the fact that the periodic firing pattern is irrelevant with either navigation algorithm (i.e., path integral or vision based) or network structure (i.e., LSTM or convolutional neural network) supports that grid cells can serve as a translation vector between input stimuli and output responses. Second, given that DNNs provide a potential representation for the grid cell, how can we train the network and replicate the specific firing pattern? One recent work proposed that grid patterns depend on the boundary shapes of the navigation environments[11]. Sorscher et al. investigated the influence of various constrains and activation functions[12]. Our work demonstrates that both training hyperparameters and output encoding methods have an essential impact on the formation of grid cells. As exemplified in the work by Banino et al. in[10], a smoother, hexagonal grid representation emerges only under a slower learning schedule with gradient clipping. We found that a quadrilateral grid, instead of a hexagonal grid, would appear when the learning rate is beyond a certain threshold and the gradient clipping is loose. Besides the learning rate, only with certain output encoding mechanisms will the grid-like representations arise. Third, does the grid firing pattern also apply to DNNs on non-spatial cognition tasks? Grid cells are not just a representation of space. Recent researches stressed the important role played by grid cell in the concept space[4,5] to map concepts to a simplified two-dimensional space. Another research reported the reuse of grid cells to encode frequency information and abstract task information[13]. In addition, grid cells are increasingly considered as enabling a translation vector model which encodes generalized and abstraction knowledge as translation vectors[14]. Bicanski et al. proposed a model to explain how human beings use saccades to explore and identify familiar objects by encoding the input-to-output association (i.e., translation) with grid cells[14]. If DNNs have a fundamental connection with biological neural networks, it is appealing to study if the grid firing pattern appears in DNNs designed for non-spatial tasks. To the best of the authors' knowledge, this is the first work establishing that non-grid spot-like firing patterns can be attained in a deep neural networks for non-spatial cognition tasks with the popular CIFAR[15] and MNIST[16] datasets. With such a non-spatial experiment, we further confirm that neural encoding mechanism is critical to the formation of grid-like representations. Our results suggest that grid cells serve as a universal neural representation across a wide variety of tasks.

# Results

We conducted four sets of experiments. First, we built a memory-less feedforward convolutional neural network (CNN), which successfully reproduced grid cells when conducting pure vision-based pathfinding. Second, we extended the aforementioned CNN by concatenating it with an LSTM to perform vision-based navigation task. Third, we evaluated the impact of hyperparameters and encoding mechanisms in a LSTM based navigation framework [10]. Finally, we proceeded to non-spatial tasks and proved that grid-like firing pattern could also be observed. To analyze experimental results in a unified set up, all firing patterns and spatial correlations in this work are plotted using the method that was used by Banino et al. in[10].

*Grid firing behaviors of DNNs with regard to training hyperparameters and input data encoding.* A large body of work has been dedicated on studying how various parameters and input characteristics affect the performance of DNNs[17,18], but a detailed evaluation of the impacts on the grid behavior is still missing. Accordingly, we performed extensive experiments on a navigation DNN similar to the one used by Banino et al.[10] to identify the essential hyperparameters and input encoding for grid cell formation. We discovered that the most relevant hyperparameters are the learning rate and gradient clipping threshold. We tested four scenarios corresponding to the four combinations of (high, low) learning rate and (strict, loose) gradient clipping. As shown in Fig. 1A, only the setting with low learning rate and strict gradient clipping leads to the hexagonal grids. When the learning rate is relatively high, quadrilateral grid cells start to emerge. If we both remove the gradient clipping and increase the learning rate, cells with quadrilateral gird cells become pervasive. The convergence curves of the four cases are plotted in Fig. 1B to evaluate the influence of training hyperparameters on convergence. When comparing the grid cell result with the convergence curve, it can be discovered that a smooth convergence process will lead to a hexagonal grid and a more radical convergence process will lead to a quadrilateral grid. In terms of input characteristics, we use ensemble coding for the output neurons to emulate the place cells in animal brains, but one recent work showed that the data encoding is not critical to the formation of grid cells[11]. We also investigated the situation when only partial navigation space was encoded by place cells. The distribution of place cells is shown in Fig. 1C. In the firing maps shown in Fig. 1D, no apparent grid structure can be identified if place cells can only be activated within a certain region of the navigation environment in spite of the fact that the model converges to a very low level of loss.

*A memory-less network for pure vision based navigation.* Animals typically utilize visual signals to refine their navigation precision[19]. A CNN network was also used in the vector navigation experiments of Banino et al. as a correction mechanism for an LSTM network performing path integration[10]. It is thus appealing to investigate whether the emergent grid firing is associated with path integration only or more pervasive in all DNNs designed for spatial navigation. We devised a CNN based memory-less feedforward neural network, which exploits the vision of the agent moving in the environment as input and predicts the agent's location encoded by a groups of place cells. The network takes colored images with a size of 64x64 as input and has its output as positions encoded with place cells. Example of the 64x64 input image is shown in Fig. 2A. The training loss is the K-L divergence between the place cell encoding of target position and predicted position. Detailed information about our DNN model is elaborated in the supplemental material. The visual input dataset is generated by an artificial agent wandering in an empty square room in Deepmind's simulation environment[20]. Each sample in the dataset is the visual snapshots in the environment and the positional information of a path with a length of 200 time steps. In our experiments input data samples are shuffled to make the input more uniform. As shown in Fig. 2B, the memory-less model does exhibit the grid-like firing map, even though it does not have any recurrent connections that store past events. The emergence of grid cells further confirms that grid-like representations can be regarded as an efficient encoding mechanism for spatial information.

*A hybrid CNN + LSTM network for navigation through vision guided path integration.* To further the exploration of how visual signals lead to grid coding, we designed a hybrid navigation network by integrating the CNN introduced earlier with a path integration LSTM for vision guided navigation. Without any pre-training with labeled images, the CNN model was trained together with the LSTM model. Again, we used the dataset used previous in pure vision based navigation experiment. In the training phase, each training sample is a continual path with a length of 100 time steps starting from a random spot, while the evaluating phase uses the whole path consisting of 200 time steps. The resultant grid firing patterns are shown in Fig. 3A. As both results from both neuroscience and AI communities support that visual information helps refine the self-positioning[10,19], we devised an experiment to evaluate the impact of such refinement. In this experiment, we concatenated two individual paths, A and B, to form a path of 200 timesteps. In the new path, the visual inputs for the first 100 time steps are from path A and those for the last 100 time steps are from path B. To extract the precise position from the output encoding, we designed another multilayer perceptron (MLP) model and trained it by providing the ensembled code as input and the original position as target. The convergence curve is shown in Fig. 3B, indicating that after around 15 epochs the model is capable of extracting the positions from the coding of place cells. One sample of a concatenated path and corresponding navigation result is shown in Fig. 3C, which we may discover that the predicted position follows the path and swiftly switch to the new path with some interval points. In Fig. 3D, we compared the convergence performance of the velocity-based path integration model by Banino et al. and ours in terms of the learning curves. To make an equal comparison, we reduce the size of the dataset used in the velocity integration model such that both models are trained with an equal number of samples. The final results demonstrate that our model eventually converges to a lower level of loss at a faster convergence speed.

*Grid cell in DNNs for a non-spatial cognition task.* While neuroscientists offered evidence that human could reuse the grid cells as an abstract map for non-spatial tasks[4,5,13,14], there are no direct evidence supporting that DNN can accomplish the same feat. We designed a non-spatial cognition experiment with the MNIST[16] dataset and CIFAR10[15] dataset and seek if the grid cell characteristics can be discovered in non-spatial tasks. As the grid cells are inherently associated with a 2-D plane, we recast the image classification problem by treating it as a 2-D mapping problem on an abstract cognitive map. In the experiment designed by Constantinescu et al.[4], the input, i.e., lengths of bird neck and leg, can be straightforwardly mapped to a 2-D Cartesian plane encoding the concept space. Such a natural mapping turns out to be hard to find when using perceptual data with high-dimensional features. In this work, we use T-SNE[21] to map images in a high dimensional space to data points in a 2-D Cartesian plane. While the MNIST samples are directly mapped onto the plane, the samples from CIFAR10 are first fed to a pre-trained classification network (ResNet18) without the last output layer and the extracted features are then mapped to a plane with T-SNE. The spatial distribution of the two datasets are shown in Fig 4A, with each color corresponding to a unique category. The deep learning model for predicting the position encoded by neural ensembles is a 4 layer MLP that is identical to the network used in the spatial tasks. The number of neurons in the output layer equals the number of place cells, i.e., 256 in our experiments. The input layer takes a 784-dimensional vector for MNIST and a 512-dimensional vector for CIFAR10. The results are shown in Fig 4B. Although the firing pattern does not form typical hexagonal grids, it does exhibit firing spots in discrete locations. The results suggest that grid cells will fire over different input features, which may belong to different categories sharing similar features. As the experiments is not carried out on a dataset with typical planar distribution, we raise the hypothesis that grid cells actually serves as a universal representation in both spatial and non-spatial cognition tasks.

# Discussion

Hippocampus and its closely connected brain regions like medial entorhinal cortex have long been speculated as a cognitive map, which stores prior knowledge and associative memory for the purpose of abstraction and generalization, to sustain intelligent behaviors. The discovery of grid cells and other navigation related cells significantly boost our understanding towards how brains perform navigation tasks. The discovery of the emergence of grid cells in artificial neural networks suggests that such cells are playing a more fundamental role in neural network based navigation tasks [10]. Our work further pushes forward the understanding of grid cells and at the same time poses new questions.

Our experiments on vision based and hybrid vision and path integration based navigation, together with results on the encoding mechanisms and velocity-based path integration [10,11], prove that grid cells are ubiquitous in artificial neural networks designed for navigation tasks. Such an observation justifies the speculation that grid cells can serve as an effective spatial representation that is independent with sensory inputs and neural implementations. In other words, the neural networks, ether biologically plausible spiking neural networks or artificial neural networks, are capable of understanding and transferring the input sensory stimuli to the position represented by grid cells. Moreover, we show that grid cells can be reproduced in DNNs designed for a non-spatial cognition task, i.e., image recognition, through a reformulation the original problem by mapping categories to a planar organization. As the development of spatial cognition capabilities was much earlier than that of non-spatial capabilities, we human beings (and other animals with such capabilities) probably reuse the spatial representation to resolve other cognition tasks. Grid cells then provide a bridge between sensory and memory systems in hippocampus, which governs the organization of priors and abstract representations for various cognition activities[22,23]. Our experimental results offer evidence that a set of grid cells can serve as the representation of abstract knowledge in an artificial neural network. Modern DNNs are widely criticized as lacking the capability of reasoning with common sense[24]. We speculative that grid cells can serve as a long-missing cognitive map for DNNs to support human-level generalization and commonsense based reasoning.

While we are optimistic about the potential of DNNs for processing abstract knowledge and associative reasoning, our results elicit a series of questions worth further investigations.

Our experiments show that the successful formation of grid cells in DNNs depends on a few training parameters and input encoding. The exact cause of such behaviors and how to measure the optimality of grid cell encoding, however, are not clear yet. As a result, manual fine-tuning is still necessary to train networks with a hexagonal grid. We speculate that a similar set of constraints also apply to the biological neural networks and expect experimental confirmation.

So far grid cells have been discovered in animal that navigate in a largely 2-dimensional space. Existing neuroscience studies on mapping objects onto a 2-D space (e.g.,[4,5]) only handled highly simplified cases. On the one hand, it is intriguing to study if grid-like code of a higher dimension can encode more complex relationships to support a higher level of cognitive capabilities. On the other hand, how such a 2-D map encodes non-spatial relations, as objects in real world tend to be perceived as a high-dimensional feature vector, remains to be an open question. Recently researchers follow the consideration that the mapping of high-dimensional features to the 2-D grid is likely to be purely associative, i.e., disregarding the similarity between features, while the grid cells serve as a translation vector to represent the associative relations. Our finding offers evidence to this line of thinking as our model has learned to organize an arbitrary conceptual mapping.

The emergence of grid cells is likely the reflection of regularity inherent in the underlying cognition task. Recent works already show that grid cells can encode latent states in biological neural networks for reinforcement learning tasks[25]. Our finding of the grid-like representations in DNNs suggests a new path toward leveraging such a universal representation to enable abstraction and generalization in artificial neural networks.

# Methods

**Positional information encoding**

We encode the place information by adopting the approach proposed by Banino et al.[10]. A total of 256 random points are generated in the experiment room. To encode a position in the room, the encoding of a place cell follows a Gaussian probability with a given standard deviation and is normalized by dividing the summation of probabilities. Suppose $(x, y)$ is the position to be encoded with the ensemble coding place cells and $(x_i, y_i), i = 1,2, \ldots 256$ are random generated place cells, $\sigma$ as the standard deviation of Gaussian of each place cell, the encoded value of a place cell w.r.t $(x, y)$ as:

$$P_i = \frac{\exp\left(-\frac{(x_i - x)^2 + (y_i - y)^2}{2\sigma^2}\right)}{\sum_j \exp\left(-\frac{(x_j - x)^2 + (y_j - y)^2}{2\sigma^2}\right)}$$

The head direction information is also encoded by 12 head direction cells (1), but is not the major concern of this paper.

**Extracting positions from place cells**

To extract the positions from the ensemble coding of place cells, we trained another multilayer perceptron (MLP). Within each batch of training, a given number, i.e., a batch size of 16 in our experiments, of positions are generated and then encoded with the ensemble mechanism of place cells. Then the codes are fed to the MLP as input and the corresponding positions are provided as the output target of the MLP with the loss measured with mean square error (MSE). Each training epoch consists of 1000 training batches. In about 15 epochs, the model converges to a level of loss that is capable of extracting the positions from the place cell codes.

**Generation of our vision dataset**

The vision dataset is generated by recording the vision images and position information provided by the Deepmind's Lab[20] environment. A path of an agent is generated by setting a random target and having the agent to chase it by performing such actions as turning, speeding up, and slowing down. The agent is spawn from the center of the room. After a certain number (40 in our experiments) time steps, the target is changed and the agent starts to chase the new target. Such a strategy is to make a path more reasonable, i.e., not simply always going straight.

**Mapping of non-spatial categories**

The non-spatial categories of datasets used in our experiments, i.e. MNIST and CIFAR dataset, are mapped onto 2-D plane by T-SNE[21]. The samples in MNIST are directly fed to T-SNE and mapped onto the plane. In the case of CIFAR, a 512-dimensional feature for a sample is first extracted by a pre-trained neural network and then passed to T-SNE. The feature then serves as the input to the deep neural networks for the non-spatial task on CIFAR10. As T-SNE produces some outliers for every category, the isolation forest algorithm[26] is adopted for outlier removal. The contamination value is set to be 0.1 to rule out 10% of total points in one category.

**Model Specifications**

In our experiments, we designed three models of deep neural networks, a memory-less convolutional neural network (CNN) for navigation, a hybrid LSTM for navigation, and a CNN for a non-spatial cognition task (i.e., image classification). The network models are illustrated in Fig. 5. The models used for navigation tasks take image as shown in Fig. 2A as input. The output target positions are encoded with 256 place cells and 12 head direction cells as described before. Each convolutional layer is followed by a batch normalization layer, which is not illustrated in Fig. 5 for simplicity. All activation functions in the CNN frontend are ReLU, while no activation is applied to the succeeding layers in the memory-less model and the hybrid model except for those explicitly designated in the LSTM cell. The non-spatial model is a 4-layer MLP. Each layer has 256 units, and the input layer takes vector with length of 784 for MNIST[16] and vector with length of 512 for CIFAR[15] features.

**Measure of gridness**

Gridness reflects the spatial regularity of grid cells. We adopted the gridness measure used by Banino et al.[10], which was introduced by Sargolini et al.[5]. To evaluate the gridness, the spatial autocorrelogram is calculated by evaluating the Pearson product of the original heatmap. The spatial autocorrelogram at point $(x_\tau, y_\tau)$ is derived as:
$$S_\theta(x_\tau, y_\tau) = P(f(x,y), f_\theta(x - x_\tau, y - y_\tau))$$
where $f(x, y)$ is the value at point $(x, y)$ of heatmap and $f_\theta(x, y)$ is the value of $(x, y)$ of heatmap that was rotated for angel $\theta$. And the Pearson product is defined as:
$$P(x, y) = \frac{n\sum x_i y_i - \sum x_i \sum y_i}{\sqrt{n\sum x_i^2 - (\sum x_i)^2}\sqrt{n\sum y_i^2 - (\sum y_i)^2}}$$
Then the spatial autocorrelogram is rotated by a certain angle to calculate the correlation coefficient between the rotated and original autocorrelograms. For hexagonal grids, the difference between the mean of correlation coefficient of rotating the spatial autocorrelogram by 60 degrees and 120 degrees and the mean of rotating it by 30, 90 and 150 degrees is computed as a measure, because a hexagonal grid expects the greatest correlation when rotating by 60 and 120 degrees and the lowest correlation when rotating by 30, 90, and 150 degrees of rotation. And for quadrilateral grids, the difference between the mean of correlation coefficient with rotating 90 degrees and the mean of rotating the autocorrelograms by 45 and 135 degrees is computed.

$$r_\theta = S_\theta(0,0)$$
$$hexagonal\ gridness\ score = \frac{r_{60°} + r_{120°}}{2} - \frac{r_{30°} + r_{90°} + r_{150°}}{3}$$
$$quadrilateral\ gridness\ score = \frac{r_{90°} + r_{180°}}{2} - \frac{r_{45°} + r_{135°}}{2}$$

A ring mask is used to control the area of spatial autocorrelogram evaluation. The mask has an initial inner radius of 0.2, while the outer radius is increased from 0.4 to 1 with a step of 0.06. To determine the best scale of gridding, only the ring corresponding to the best gridness is kept.

**Detailed experiment settings**

*Grid firing behaviors of DNNs with regard to training hyperparameters and input data encoding.* This experiment is based on the open source code provided by Banino et al.[10]. The training hyperparameters experiments consist of four scenarios corresponding to the four combinations of (high, low) learning rate and (strict, loose) gradient clipping. The high learning rate is set as 1e-3, while the low learning rate is 1e-5. Strict gradient clipping comes with a clipping threshold of 1e-5 and the loose one a clipping threshold of 1. In the input data encoding experiments, we modified the distribution region of place cells by limiting them to a half, a quarter, one tenth, and the top right corner of the entire region.

*A memory-less network for pure vision based navigation.* The model used in this experiment is described in the methods. The input to the model is an image from our vision dataset as described in Materials and Methods. The output target is the corresponding position of the input image. The images and positions of one batch are random selected from a path with a length of 200 steps. The training optimizer is a RMSprop[27] with a learning rate of 1e-5 and a momentum of 0.9. Gradient clipping with a threshold of 1e-5 is only applied to the backend predictor illustrated in Fig. 5.

*A hybrid CNN + LSTM network for navigation through vision guided path integration.* The model used in this experiment is described in Materials and Methods. The input of this model is a sequence of images along a path. Each path sample in the dataset has a length of 200 continuous steps. In the training stage, only 100 continuous steps starting from a random point are used to for better generalization. In the evaluation stage, the full path is applied for simplicity. For each sample in a batch, we used a different path instead of a different starting point to increase the variance within the batch data. The training parameters are chosen to be similar to those of the memory-less network. Gradient clipping is only applied to the backend predictor illustrated in Fig 5.

*Grid cell in DNNs for a non-spatial cognition task.* The model used in this experiment is described in Materials and Methods. The input of this model is a 32x32 3-channel RGB image in CIFAR10 dataset[15] and the output target is the mapped coordinates, i.e., the categorization result, of the input image. The training mechanism is slightly different from those used before. We use a Adagrad[28] optimizer for stacked CNN blocks in ResNets. We set the learning rate as 1e-3 and no gradient clipping is applied to the model.

## Code Availability

All codes are available from the corresponding upon request.

## Data Availability

Our vision dataset is too large, so we suggest to generate your own dataset with our code and we are willing to provide the corresponding codes.

# Figures

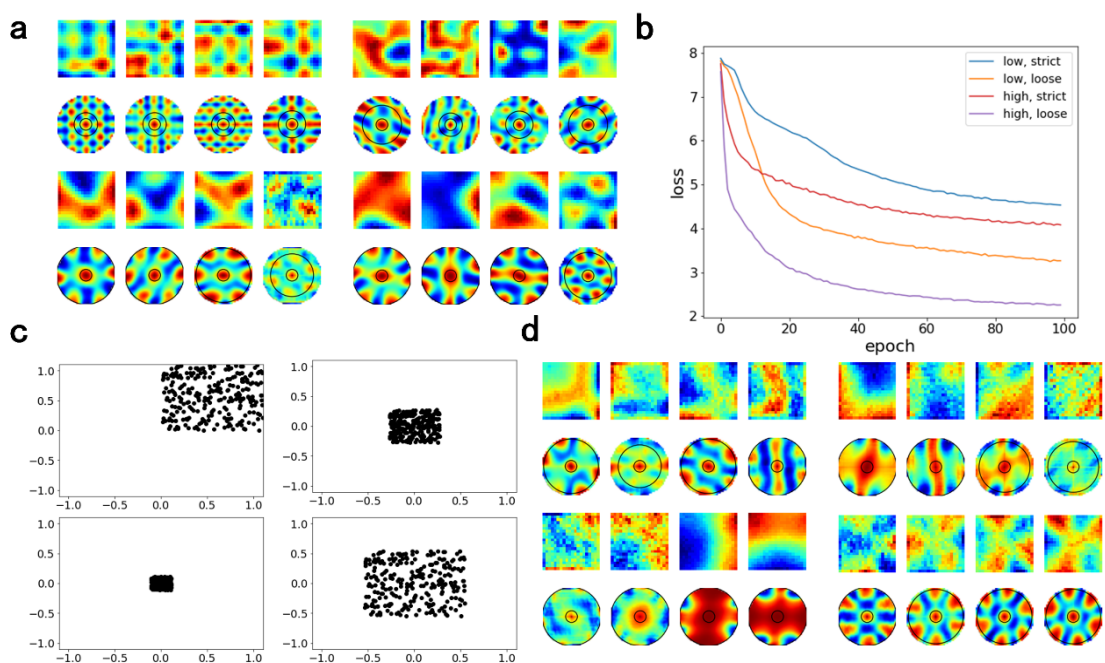

**Fig. 1** Convergence behaviors of grid cells in Banino et al.'s work under varying settings of training parameters and encoding. **a** Convergence results under varying settings of training parameters. Each setting has two rows and four columns. The first row is the activation heatmap and the second row is the spatial autocorrelograms. Each column corresponds to a cell. Top left: high learning rate and loose gradient clipping, top right: high learning rate and strict gradient clipping, bottom left: low learning rate and loose gradient clipping, bottom right: low learning rate and strict gradient clipping. All heatmaps and spatial autocorrelograms are plotted using the code by Banino et al[10]. **b** Convergence curves for all the four settings mentioned in **a**. **c** Distribution of place cells' location in different experiment settings. d Grid behaviors for different place cells' locations corresponding to those in **c**.

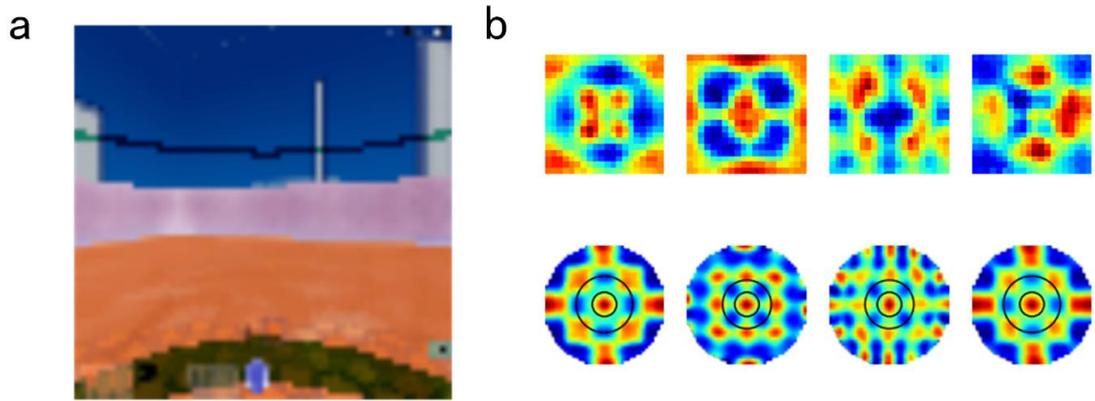

**Fig. 2 Results of memory-less navigation DNN. a** An example of visual input in the generated dataset. **b** Grid-like behaviors of the memory-less **DNN**.

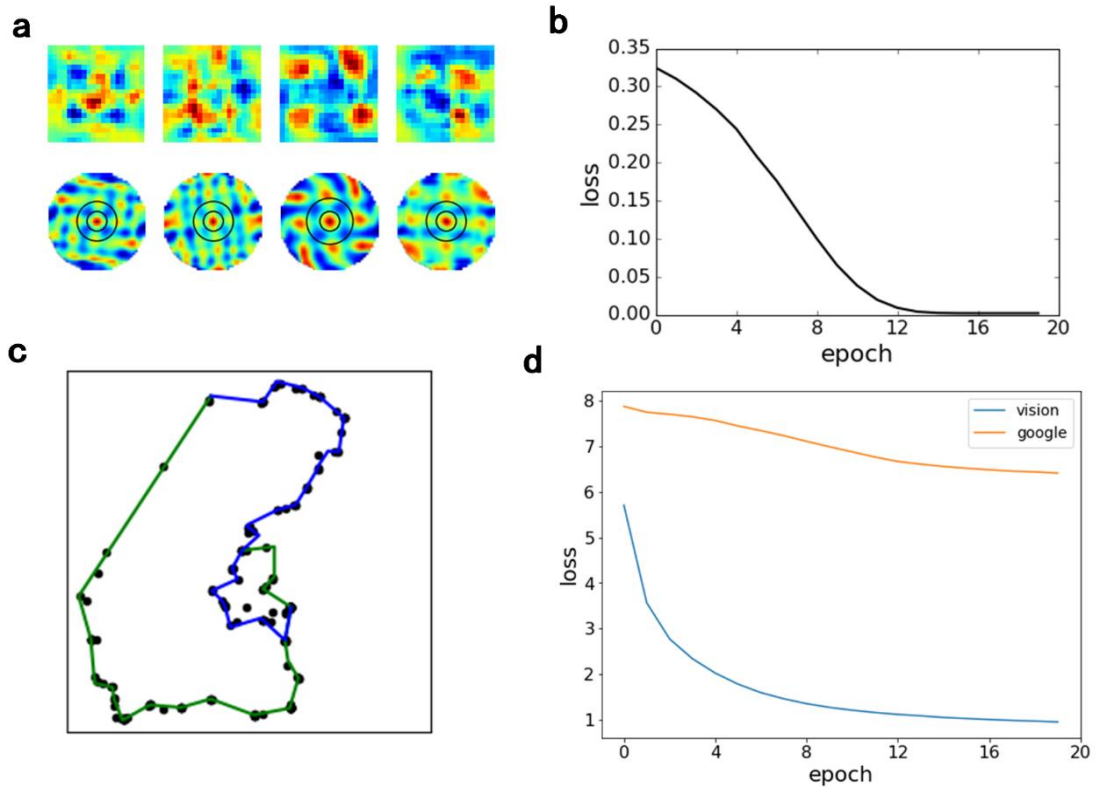

**Fig. 3** Results of the hybrid CNN+LSTM navigation model. **a** Gridding result of the hybrid network. Quadrilateral grid cells can be observed in this model. **b** The convergence curve for the position extraction model. The y-axis is the MSE loss between predicted locations by the model from the ensembled code and the actual ensembled location. **c** An example of concatenated path, the green line represents for the first 100 steps, and the blue line represents for the last 100 steps. The black dots are the locations extracted from the encoding of output place cells. The result shows that the hybrid model can swiftly adapt to the change. **d** Comparison of model convergence speed between hybrid model and the LSTM designed by Banino et al. Our hybrid model converges significantly faster to a lower level of error.

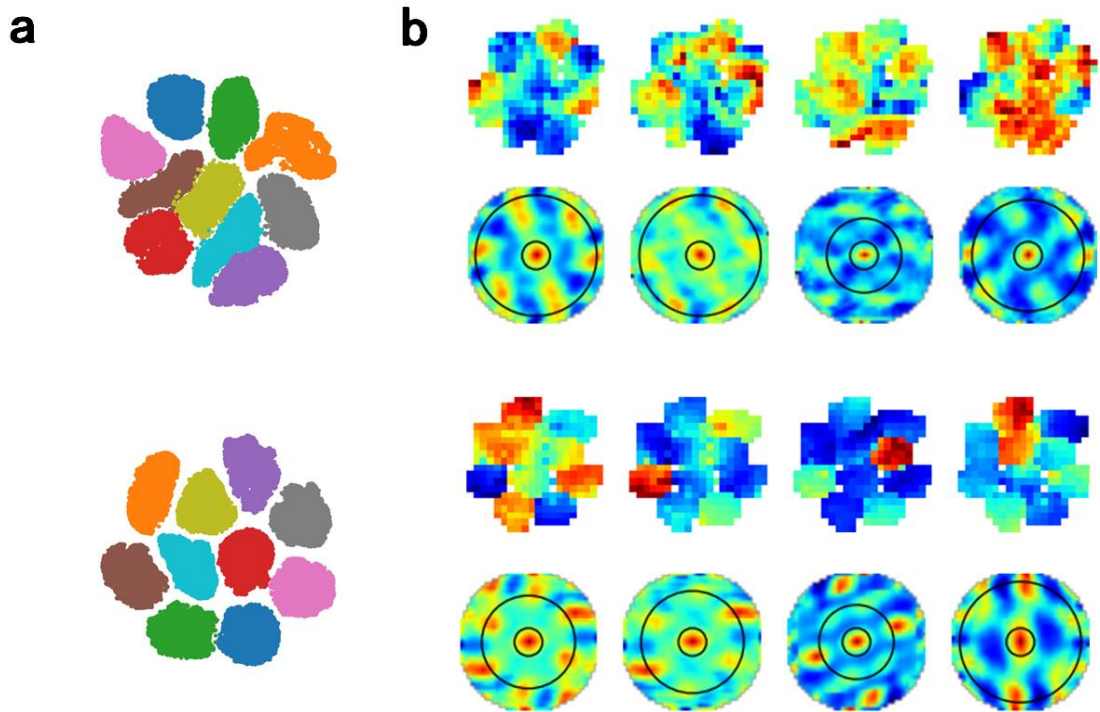

**Fig. 4** Grid-like behavior in a non-spatial cognition task. **a** Distributions of categories of different datasets Each color represents one category. The upper figure corresponds to the categories in MNIST and the bottom figure corresponds to CIFAR10. **b** Activation heatmaps and spatial autocorrelograms of the cognitive model run on the two datasets mentioned above. The cells exhibit non-grid spot-like firing patterns on both datasets.

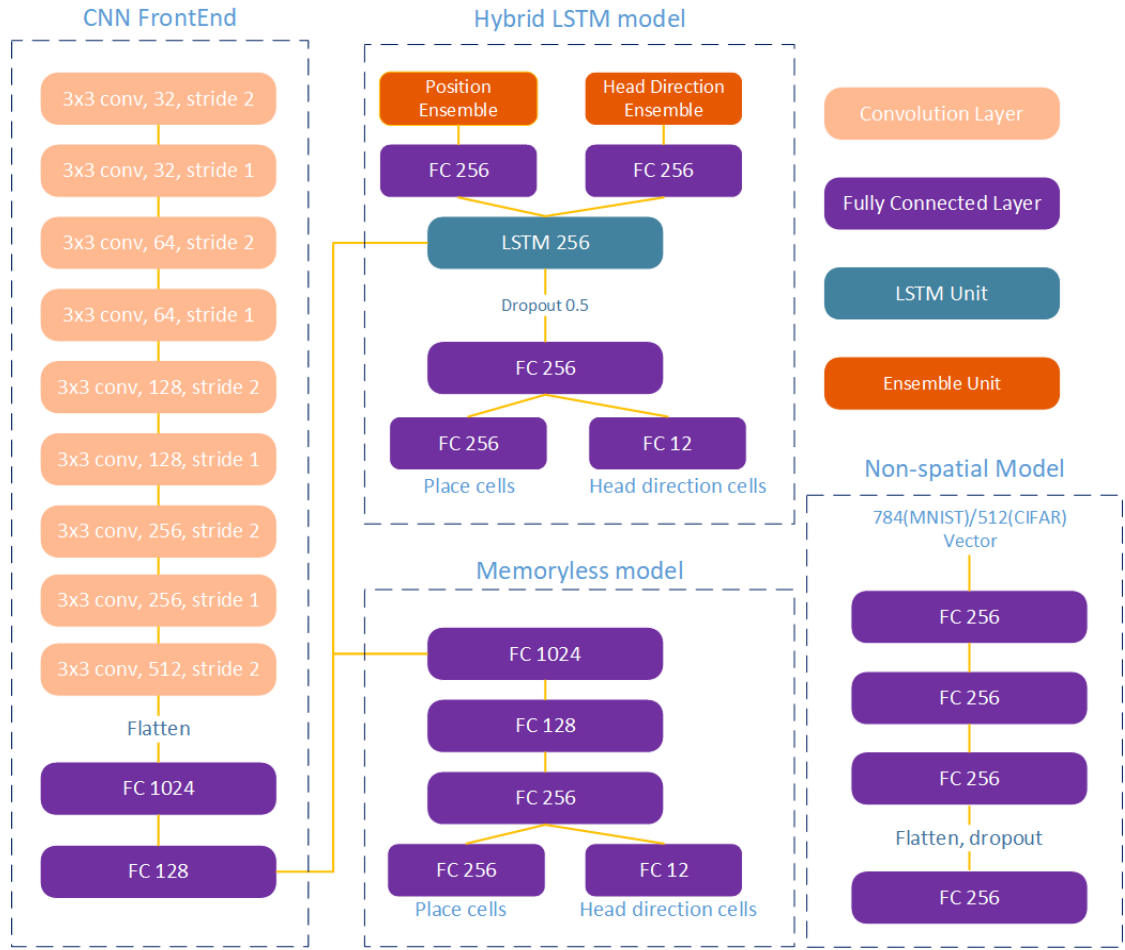

**Fig. 5** Schematic diagram for all three models used in our experiments. Each dash box represents one model. Especially, hybrid model and memoryless model uses the same CNN frontend network, but do not share the weights. The blocks are represented by different colors as illustrated in the middle left of the figure. The bottom right colored boxes are detailed illustrations of building blocks of non-spatial model, which is based on ResNet's model design[29] on CIFAR[15] dataset.